\documentclass[12pt,paper,a4paper]{JHEP3} 

\usepackage{graphicx}

\usepackage{epsfig,multicol,bbm}

\newcommand\fverb{\setbox\fverbbox=\hbox\bgroup\verb}
\newcommand\fverbdo{\egroup\medskip\noindent%
			\fbox{\unhbox\fverbbox}\ }
\newcommand\fverbit{\egroup\item[\fbox{\unhbox\fverbbox}]}
\newbox\fverbbox

\def\hxi{{\hat \xi}}

\def\N{{\mathcal N}}
\def\bx{{\bf x}}
\def\Mq{M}
\def\Mqo{M_o}

\def\E{{\mathcal{E}}}
\def\xb{{\bar x_b}}
\def\Zp{{\mathcal{Z}}}
\def\Ds{{\mathcal{D}_s}}
\def\D{{\mathcal{D}}}
\def\sym{{\rm sym} }
\def\dd{{\rm d}}
\def\mred#1{{#1}}

\def\lssq{{\ell_s^2}}
\def\t{{\hat t}}
\def\z{{\hat z}}
\def\x{{\hat x}}

\def\X{{\hat X}}
\def\w{{\hat \omega}}
\def\bepsilon{{\hat \epsilon}}
\def\n{{\hat{n}}}
\def\bp{{\bf p }}
\def\zm{{z_-}}
\def\zp{{z_+}}
\def\zpm{{z_\pm}}

\def\Fw{{F_\omega}}
\def\Fcw{{F^*_\omega}}
\def\Fwr{{\mathcal{F}_\omega}}

\def\b{{b}}
\def\wsh{{w}}
\def\Tenwsh{{T_o(z_w)}}
\def\Tenp{{T_o(\z_+)}}
\def\Tenm{{T_o(\z_-)}}
\def\Tenpm{{T_o(\z_\pm)}}
\def\Tenb{{T_o(z_b)}}
\def\Ten{{T_o}}
\def\llangle{\left\langle}
\def\rrangle{\right\rangle}

\def\F{{\mathcal{F}}}

\def\T{{\mathcal{T}}}

\def\be{\begin{eqnarray}}
\def\ee{\end{eqnarray}}
\def\del{{\partial}}

\def\G{{\hat G}}

\def\Eq#1{Eq.~(\ref{#1})}
\def\Fig#1{Fig.~\ref{#1}}
\def\Sec#1{section~\ref{#1}}
\def\Refs#1{Refs.~\cite{#1}}
\def\Ref#1{Ref.~\cite{#1}}

\title{Stochastic String Motion Above and Below the World Sheet Horizon}

\author{Jorge Casalderrey-Solana\\
    Physics Department, 
    Theory Unit, CERN,
    CH-1211 Gen\`eve 23, Switzerland
    \\
	E-mail: \email{jorge.casalderrey@cern.ch}
    }%
\author{Keun-Young Kim
\\
    Department of Physics \& Astronomy,
    SUNY at Stony Brook,
    Stony Brook, New York 11764, USA\\
        	E-mail: \email{keykim@ic.sunysb.edu}
    }
\author{Derek Teaney \\
    Department of Physics \& Astronomy,
    SUNY at Stony Brook,
    Stony Brook, New York 11764, USA
    \\
        	E-mail: \email{derek.teaney@stonybrook.edu}
    }%

\preprint{CERN-PH-TH/2009-135}

\abstract{
We study the stochastic motion of a relativistic trailing string 
in \mred{black hole  ${\rm AdS}_5$. } 
The classical  string solution
develops a world-sheet horizon and we determine  
the associated Hawking radiation spectrum. The emitted radiation causes
 fluctuations on the string both above and below the world-sheet horizon.
 In contrast to standard 
black hole physics, the fluctuations below the  horizon are causally connected with the boundary of AdS. We derive a bulk stochastic equation of 
motion for the  dual string and use the 
AdS/CFT correspondence to determine
the  evolution of a  \mred{fast} heavy quark in \mred{the} strongly coupled $\N=4$ plasma.
We find that the kinetic mass of the quark decreases by $\Delta M=-\sqrt{\gamma \lambda}T/2$ 
while the correlation time of world sheet fluctuations  increases by $\sqrt{\gamma}$.
}

\received{\today} 		
\accepted{\today}		


\begin{document}

\section{Introduction}
In recent years the dynamics of strongly coupled non-abelian plasmas  has been
investigated vigorously \cite{Schaefer:2009dj}.  
This interest in plasma physics was motivated in part by the heavy ion program at
RHIC \cite{Adams:2005dq,Adcox:2004mh,PHOBOSWhite}  
and the upcoming program at the LHC.  For the experimentally accessible
temperature range, the medium is close to the deconfinement transition and
the QCD coupling constant is not small \cite{TeaneyQGP4}.  Thus it is important to compare 
perturbative expectations for the QCD plasma to all available strong coupling results.
 
The AdS/CFT correspondence relates $\N=4$ Super Yang Mills (SYM) at strong
coupling and large $N_c$  to  type IIB supergravity on a curved background,
${\rm AdS}_5\times S_5$ \cite{Maldacena:1997re,Gubser:1998bc,Witten:1998qj}.  
Although $\N=4$ SYM is not QCD, the AdS/CFT
correspondence has provided new insight into  the strongly coupled 
regime and the RHIC experiments \cite{GubserReview1,GubserReview2}. 
In particular, the calculation of $\eta/s$ in $\N=4$ \cite{Policastro:2001yc,Kovtun:2004de,Buchel:2003tz},
 provided 
a concrete example of a strongly coupled plasma which  
realizes the small shear viscosity needed to explain the elliptic
flow observed at RHIC \cite{TeaneyQGP4}.
Since this work on shear viscosity many other 
transport properties of strongly 
coupled plasmas have been computed using the
correspondence. 
Of particular relevance
to this work  is the energy loss of heavy  quarks
\cite{Herzog:2006gh,Gubser:2006bz,CasalderreySolana:2006rq,Gubser:2006nz,CasalderreySolana:2007qw,Dominguez:2008vd,Giecold:2009cg}.

The steady state motion  of a heavy  quark 
traversing the  $\N=4$ plasma 
under the influence  of an external electric  field  
is represented in the gravitational theory 
by a semi-classical  trailing string \cite{Gubser:2006bz,Herzog:2006gh} -- see \Fig{trailing}. 
The momentum carried down the string 
determines the energy loss of the heavy quark  in the field theory 
\be
\label{drageq}
\frac{dp}{dt}=-\eta \gamma v\, , \qquad  \eta \equiv \frac{1}{2} \sqrt{\lambda} \pi T^2 \, ,
\ee 
where $v$ is the velocity and $\gamma=1/\sqrt{1-v^2}$ is the 
Lorentz factor.  
The momentum broadening of the quark probe was determined 
 by
studying small fluctuations around the drag string solution 
\cite{CasalderreySolana:2006rq,CasalderreySolana:2007qw,Gubser:2006nz}. 
\mred{
There is a 
point in the bulk where the quark velocity equals and then 
exceeds the local speed of light parallel to the brane. 
At this point the dual string  develops a world-sheet event horizon (ws-horizon)   which is distinct from the black hole horizon.
The associated Hawking radiation from this ws-horizon  determines the fluctuations
on the string.} At small velocities the fluctuations computed in this way are consistent with  the drag in 
\Eq{drageq}  and the fluctuation-dissipation theorem \cite{CasalderreySolana:2006rq}.


\FIGURE[t]{
\epsfig{file=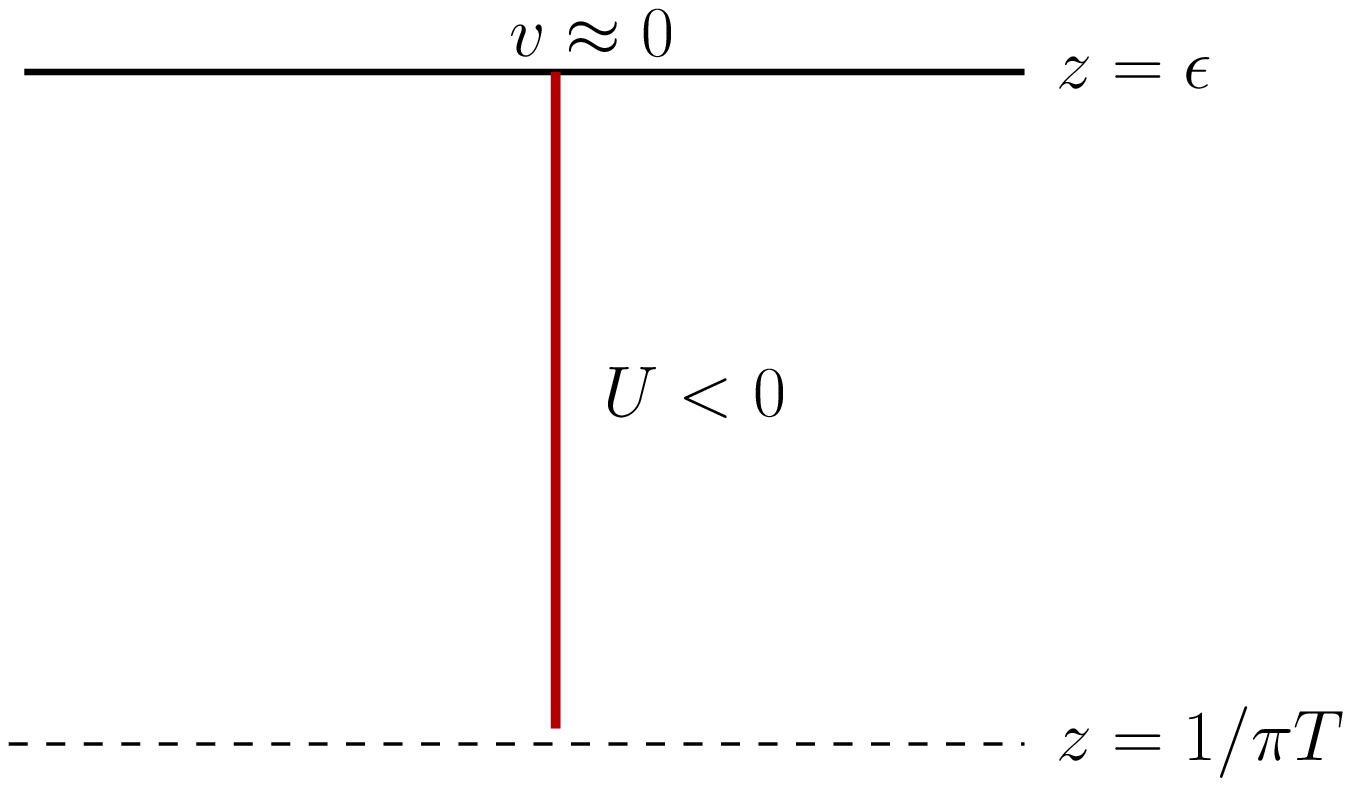,width=0.4\textwidth}
\epsfig{file=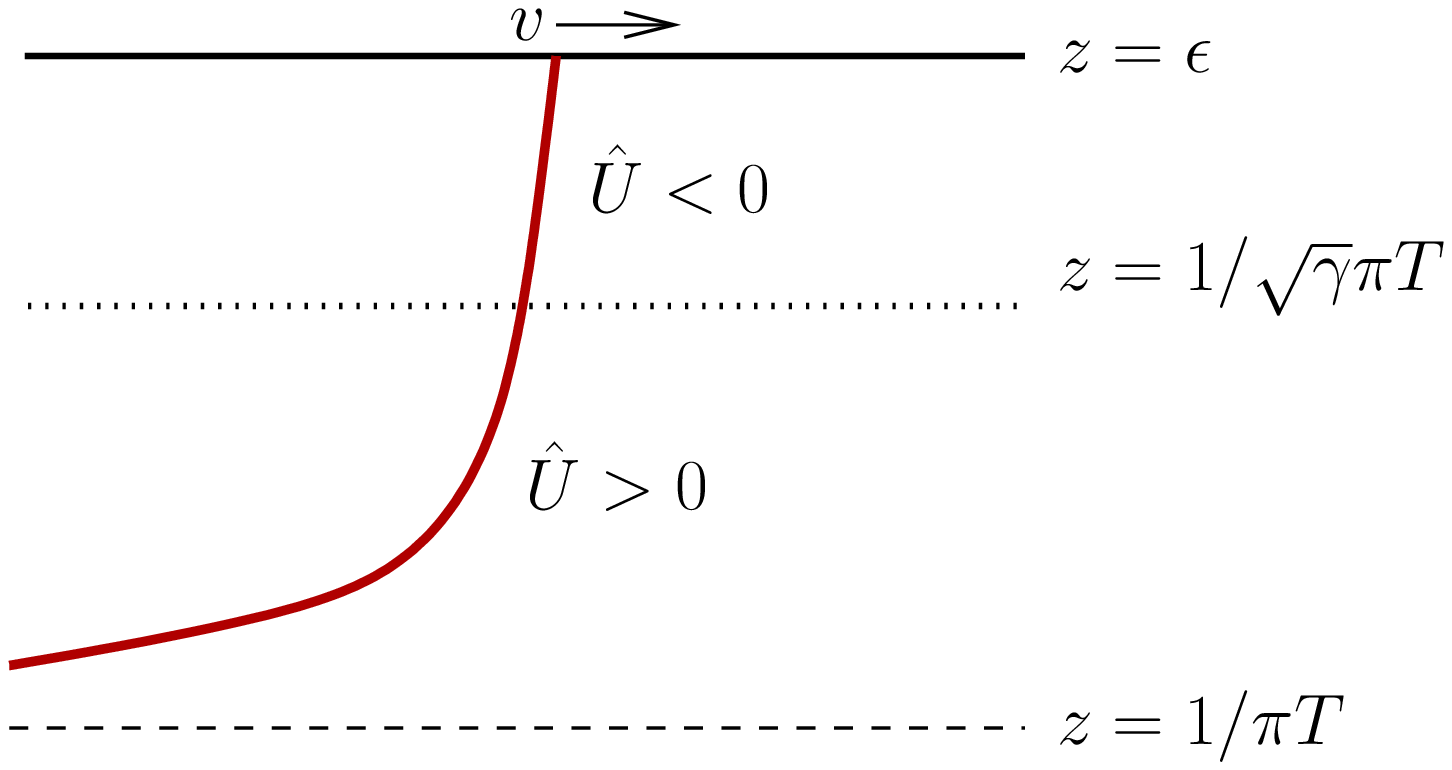,width=0.4\textwidth}
\caption 
{
(a) \mred{An approximately static heavy quark string. }
Hawking radiation induces 
a random force close to the horizon which is ultimately 
transmitted to the boundary \cite{deBoer:2008gu,Son:2009vu}. 
(b) \mred{The rapidly moving trailing string 
analyzed in this work. }
At $z=1/\sqrt{\gamma}\pi T$ the speed of the quark equals the 
local speed of light and the ws-horizon develops. Fluctuations
are induced above and below the world sheet horizon. In addition 
there is a cross correlation between the fluctuations above and 
below the ws-horizon.
}
\label{trailing}
}

The momentum broadening dicussed above was computed by  
disturbing the string with an external force, 
and subsequently following
the ring down  of this disturbance
with the classical equations of motion.
\mred{In reality external forces are unecessary since  Hawking radiation emanates from 
the  ws-horizon 
and induces
stochastic motion in the bulk.} This picture was  clarified recently 
in \Refs{deBoer:2008gu,Son:2009vu} which determined
 the  stochastic equation of motion  for a slowly
moving quark  string. (In this limit the string 
is nearly straight and  the ws-horizon
coincides with the black-hole horizon.)
The effect of  Hawking  radiation is
to supplement the dissipative boundary conditions at the
black hole event horizon with an associated random force.
 The classical string equations of motion 
transmit this random  force to the boundary, 
leading to the stochastic motion of the heavy quark. The amplitude of the random term is suppressed  by $\lambda^{1/4}$  reflecting the quantum mechanical nature of black hole irradiance.

In this work we extend the analysis of Ref.~\cite{Son:2009vu} to the rapidly moving trailing string. The random fluctuations
of the string are generated at the world-sheet horizon and propagate both upward toward the AdS boundary, and downward toward the horizon of black hole. 
Our study of the world sheet radiation determines  the statistics of 
these fluctuations both above and below the world-sheet horizon. In 
addition there is cross correlation, {\it i.e.} fluctuations above the 
ws-horizon are correlated with fluctuations below the ws-horizon.
This completes the analysis begun in \Ref{Giecold:2009cg} of the fluctuations above the 
horizon. Since the fluctuations below the horizon  are causally connected with the boundary,  
these random vibrations
will be reflected in the field theory by  the stress tensor fluctuations induced by the heavy quark.
We have also re-derived the heavy quark effective equation of motion \cite{Giecold:2009cg} and computed 
a velocity dependent mass shift.

The paper is organized as follows. 
In \Sec{review} we \mred{briefly} review the trailing string solution.  
In \Sec{sf} we compute the stochastic equation of motion of the semiclassical string in bulk by \mred{ integrating out the fluctuations in a
strip} around the world-sheet horizon.
This  bulk evolution ultimately determines
the equation of motion for string end point which is dual to the heavy quark. The heavy
quark partition function and its equation of motion \mred{are} derived in  \Sec{HQDym}. 
Finally, the \mred{bulk fluctuations are} re-derived in terms of the bulk to bulk Green\'{} functions in Appendix \ref{gfapproach}.  A discussion  of the  physical picture which emerges from this analysis is provided in \Sec{sando}.

\section{Preliminaries\label{review}}

\subsection{Finite Temperature in AdS/CFT}
\FIGURE[t]{
\epsfig{file=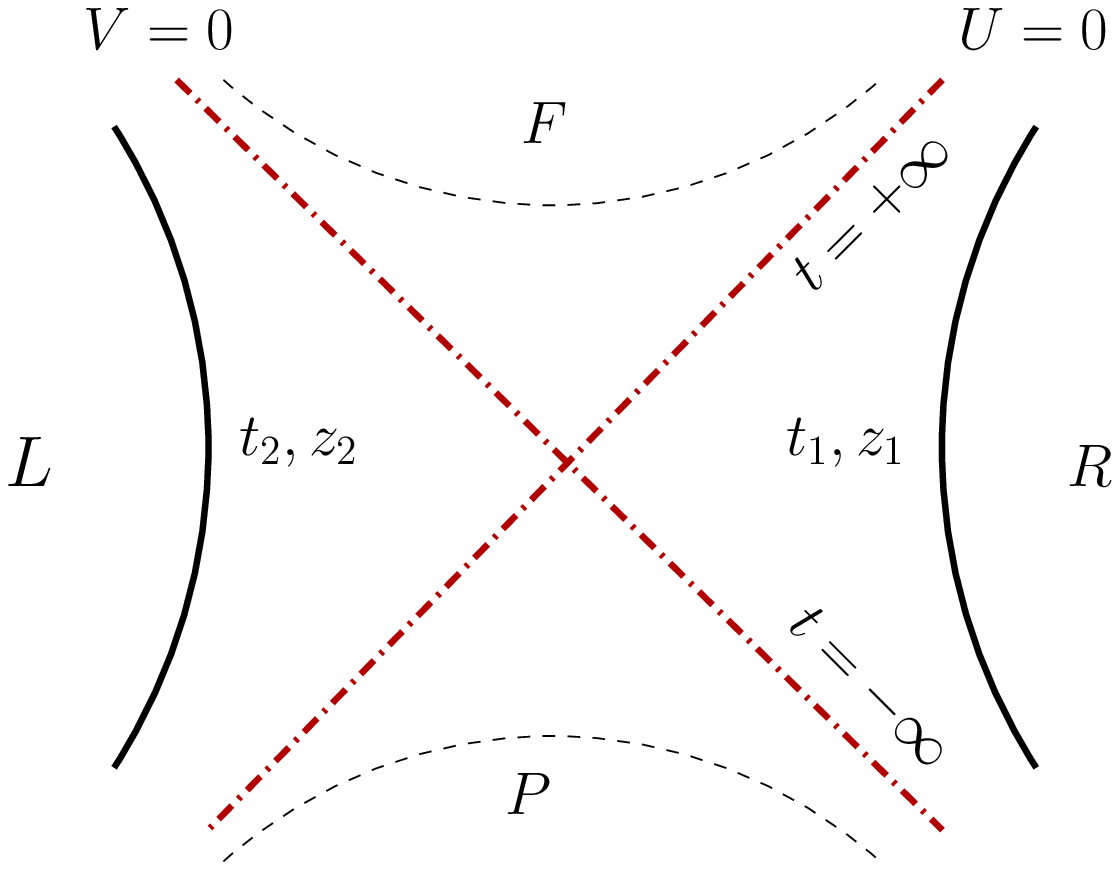}
\caption
{
\mred{Kruskal diagram for the AdS black hole. The coordinates $(t_1,z_1)$ span the 
right (R) quadrant while $(t_2,z_2)$ span the left quadrant (L). The dash-dotted lines and the dashed hyperbolas 
represent the future and past 
horizons and the singularities, respectively. The thick hyperbolas on the sides
of the two quadrants
are the
boundaries  at $r=\infty$ (or $z=0$).  }
}
\label{kruskal}
}

According to the AdS/CFT correspondence, $\N=4$ SYM at finite temperature and infinite 't Hooft coupling is dual to a black  brane solution which is asymptotically 
${\rm AdS}_5 \times S_5$.  The metric of this solution  can be written as
\be
\label{poincare}
d s^2=\frac{L^2}{z^2}
\left(
-f(z) dt^2 + d \bx^2 + \frac{1}{f(z)} d z^2
\right) \, ,
\ee
where $L$ is the AdS radius, $f(z)=1-(z \pi T)^4$, and $T$ is the Hawking temperature of the black brane. 

As is well known, the singularity of the metric at $z=1/\pi T$ in \Eq{poincare} is only a coordinate 
singularity reflecting the fact that the metric does not describe the entire space-time
of the black brane solution. A global set of coordinates, known as Kruskal coordinates, can be introduced which removes this singularity \cite{Fidkowski:2003nf}. The Kruskal map is 
shown in \Fig{kruskal}.
The set of coordinates \Eq{poincare} describes  only the right quadrant of the Kruskal plane -- see \Fig{kruskal}. An identical set
of coordinates may be introduced to describe the left quadrant of the Kruskal map. We will denote by 
\be\{t_i, x_i, z_i\}
\quad \quad {\rm with}\quad  i=1,2 \, ,
\ee
the coordinates of the right and the left quadrant respectively.  The doubling of the fields in the presence of the 
black branes corresponds \cite{Herzog:2002pc} to the appearance of type 1 and 2 fields in finite temperature field theory \cite{LeBellac}.

 \subsection{Review of the trailing string}
The dynamics of a heavy external quark in the fundamental color representation of the gauge theory are determined, 
according to the correspondence, by a semiclassical string stretching from the boundary of AdS to the horizon \cite{Maldacena:1998im}. The forced motion of the quark under a constant external electric field in the $\hat x$ direction is dual to the classical solution to the Nambu-Goto action
plus an electric field acting on the boundary endpoint
\be
S=-\frac{1}{2\pi \lssq }\int \dd t \dd z \,\sqrt{-{\rm det} 
										h_{ab}
									}
+
\int \dd t\, \E \xb(t) \, ,
\ee
where $h_{ab}$ is the induced metric, $\E$ is the external field and $\xb(t)=\bar X(t,\epsilon)$ the position of the boundary 
end point of the string.

The demand that the solution is stationary ({\it i.e.} that the action is time independent and real) fixes the velocity of the
quark in terms of the external field.
\be
v=\frac{2\E}{\sqrt{\lambda}\pi T^2 } \frac{1}{\sqrt{1+\frac{4\E^2}{\lambda \pi^2 T^4}}} \, ,  
\ee
and the string solution is given by 
\be
\label{dssol}
\bar X(t,z)=v t +v  \Delta X_{TS}(z) \, ,
\ee
with the trailing string profile \cite{Herzog:2006gh,Gubser:2006bz}
\be
\label{tsprfl}
\Delta X_{TS}(z)=
\frac{1}{2\pi T} \left[
					\tan^{-1}(z \pi T)- \tanh^{-1}(z \pi T)
					\right] \, .
\ee
In these coordinates, the induced metric on the string is non-diagonal. The induced metric is diagonalized by the set of coordinates 
\be
\label{chvrbl}
\t&=&\frac{t+\zeta(z)}{\sqrt{\gamma}} \, , 
\\
&& \zeta(z)\,\equiv\, 
\frac{1}{2  \pi T} \left(\tan^{-1}(z \pi T)-\tanh^{-1}(z\pi T)\right) 
\nonumber
\\
&& \qquad \
 -\, \frac{\sqrt{\gamma}}{2 \pi T} \left(\tan^{-1}(z\sqrt{\gamma}\pi T)-\frac{1}{2}\ln
 											\left|
												\frac{1+z\sqrt{\gamma}\pi T}{1-z \sqrt{\gamma} \pi T}
 											\right|
                    \right) 
\nonumber \, ,
\\
\z&=&\sqrt{\gamma} z \, , 
\\
\hat \bx&=&\sqrt{\gamma}\bx  \, ,
\ee

This change of variables is singular at the position $z=1/\sqrt{\gamma} \pi T$ which is the world-sheet horizon. 
The induced string metric in these coordinates is given by
\be
h_{\t\t}=-\frac{L^2}{\z^2} f(\z) \,, \quad h_{\z\z}=\frac{L^2}{\z^2}\frac{1}{f(\z)} \,, \quad h_{\t\z}=0 \, , 
\ee  
which shows a horizon at $\z=1/\pi T$. The singular change of variables \Eq{chvrbl} should be understood as two different 
local charts describing the space above and below the world-sheet horizon. 

The fluctuations around the solution \Eq{dssol} are described by the Nambu-Goto action.
 For small fluctuations it is enough to expand this action to quadratic order. Denoting by $\X_L(t,z)$ and $\X_T(t,z)$ the longitudinal and transverse fluctuations, the action is 
\be
S_2[\X_L, \X_T]&=&  S_L[\X_L] +S_T[\X_T]  \,, 
\ee
with
\be
\label{S2def}
S_T[\X]&=&-\frac{1}{2}\int \dd \t \dd\z\,
		\left[
			\left(
			\Ten(\z) \left(\del_\z \X\right)^2-\frac{m}{ \pi T \z^2 f(\z)} \left(\del_\t \X \right)^2
			\right)
		\right] \, ,
		\\
S_L[\X]&=&\gamma^2\,S_T[\X] \, ,
\ee
where the subscripts $T$, $L$ denote transverse and longitudinal fluctuations and 
\be
\label{Tensiondef}
\Ten(\z)=\frac{L^2}{2\pi\lssq} \frac{f(\z)}{\z^2}\, , \quad  m=\frac{\pi T L^2}{2\pi \lssq} \,.
\ee
We remark that the dynamics of the longitudinal and transverse fluctuations are the same and they 
only differ by an overall $\gamma^2$ factor in the action.
  The classical  equations of motion are, thus, the same for 
longitudinal and transverse modes and, in Fourier space,
are given by 
\be
\label{eomX}
\del_\z \left(\Ten(\z) \del_\z \X(\w,\z)\right)+\frac{m \w^2}{\pi T \z^2f(\z)}\X(\w,\z)=0 \,.
\ee
Classical solutions to this equation can be expressed in terms of the in-falling solution 
\be
\label{Fdef}
\Fw(\z)=\left|1-\z \,\pi T\right|^{-i\w/4\pi T}\Fwr(\z) \, ,
\ee 
with $\Fwr(\z)$ a regular function.  For arbitrary $\z$ and $\w$, the function $\Fwr(\z)$ is only known
numerically; however an analytical expression for the solution can be obtained in the 
low frequency limit
\be
\Fw(\z)=1 -\frac{i\w}{2\pi T}\left(
					\tan^{-1}(\z)- \frac{1}{2} \ln \left|
										\frac{1+\z \pi T}{ 1-\z\pi T}
										  \right|
					\right)
					+\mathcal{O}(\w^2) \, ,
\ee
which is valid as long as $z$ is not exponentially close to the world-sheet horizon
 ($\w \ln (1-\z\pi T)\ll1$). We would like to remark that, even though the $z-$derivative of this
 expression has a kink at $\z=1/\pi T$ the tension force of the string close to the horizon is the same on both sides
 of the horizon
 \be
 \label{tencont}
 \left .\Ten(\z) \del_\z \Fw(\z) \right|_{\z=(1+\epsilon)/\pi T}
\approx
 \left .\Ten(\z) \del_\z \Fw(\z) \right|_{\z=(1-\epsilon)/\pi T} \, ,
 \ee

The in-falling solution $F_\w$ determines the retarded correlator of the random force acting on 
the quark as it propagates through the strongly coupled thermal medium  \cite{CasalderreySolana:2006rq}
\be
\label{GRdef}
G_R(\w) = - \Ten(\z) F^*_\w (\z)\del_\z F_\w(\z) \, .
\ee
As noted in \cite{Son:2009vu} the imaginary part of $G_R$ coincides with the Wronskian of  \Eq{eomX} 
and, thus, it is independent of $\z$.

  A general solution of the equation of motion can be expressed as a linear combination of 
  $\Fw(z)$ and $\Fw^*(z)$. Analyticity demands 
that the solutions below and above the world sheet horizon are connected  by \cite{CasalderreySolana:2007qw}
\be
\label{clsl}
\X_1(\w,\z)=a_1(\w)e^{ \theta(\z-1)\w/2T} \Fcw(\z)+ b_1(\w) \Fw(\z) \, ,
\ee
where we have added the subscript $1$ to denote that the these string fluctuations take place in the right 
quadrant of \Fig{kruskal}. A similar analysis may be performed in the left quadrant yielding
\be
\X_i(\w,\z)=a_i(\w)e^{ \delta_i\theta(\z-1)\w/2T} \Fcw(\z)+ b_i(\w) \Fw(\z) \, ,
\ee
with $i=1,2$
and $\delta_1$=1 and $\delta_2=-1$. The fluctuations in both quadrants are also related by 
analyticity. 
Following \cite{
CasalderreySolana:2007qw}  
\be
\label{jumphorizon}
a_2(\w) = e^{-\omega \sigma}  e^{\w/T} a_1(\w) \,, \quad \quad  b_2(\w) = e^{-\omega \sigma}  b_1(\w) \,,
\ee
with $\omega=\w/\sqrt{\gamma}$ the boundary frequency in the $t$ coordinate and $\sigma=1/2T$. 

As noted in \cite{Herzog:2002pc}, the identification of $X_2$ with the type 2 fields of the
 Schwinger-Keldysh contour corresponds to the choice of $\sigma=1/2T$. And arbitrary choice
 of $\sigma$ may be found by performing the change of variables \cite{Son:2009vu}
 \be
 X^{(\sigma)}_2(\w, \z)= e^{\omega(1/2T-\sigma)} X_2(\w,\z) \, .
 \ee
With this choice, the relation \Eq{jumphorizon} still holds if we now identify $\sigma$ with the 
$\sigma$-contour. From now on we will focus on $\sigma=0$ and drop the $\sigma$ superscript. 
 
\section{String Fluctuations from the Stretched World-Sheet Horizon\label{sf}} 
\subsection{The Partition Function}
The discussion in the section \ref{review} was concentrated in the strict $\lambda\rightarrow\infty$ limit
in which  the string partition function is saturated by the classical action and the motion of the 
string is purely classical. String (quantum) fluctuations are suppressed by $\lambda$. As shown in 
\cite{Son:2009vu},
 in order to recover the stochastic motion of the heavy quark, we must take $\lambda$ large but finite
and consider the quantum fluctuations induced by the world-sheet horizon (Hawking radiation). 

Since $\lambda$ is large, we shall concentrate on small fluctuations around the local minimum
of the classical action given by the solution \Eq{dssol}. The partition function is then given by
\be
\Zp_s&=&\Zp_T \, \Zp_L \, ,
\\
\Zp_\alpha&=&\int \Ds \X_1(\t,\z) \Ds \X_2(\t,\z) 
e^{i S_\alpha[\X_1]-iS_\alpha[\X_2]} \, ,
\ee
with $\alpha=L\,,T$. The measure $\Ds$ is a complicated object which we will not need to specify. 
Since the partition function factorizes and the classical action for longitudinal fluctuations is 
proportional to the transverse ones, we will concentrate in the latter and we will drop the 
subindex $T$. The extension to longitudinal fluctuations is straightforward and we will quote
the main results at the end.

The partition function integrates over all values of $\z$ and, in particular, across the world-sheet 
horizon. Since the set of coordinates \Eq{chvrbl} do not cover the hyper-plane $\z^\wsh=1/\pi T$ we 
must integrate out the region surrounding this hyper-plane.  We introduce
\be
\x^\wsh_{i\pm}(\t)=\X(\t,\z_\pm) \, , \quad \z_\pm=(1\pm\bepsilon)\z^\wsh \,.
\ee
%
%
%
and express the partition function as
\be
\Zp&=&\int \D\x^\wsh_{1-} \D\x^\wsh_{1+} \D\x^\wsh_{2-} \D\x^\wsh_{2+}
\\ \nonumber
&& \quad \quad \quad
\Zp^{<}[\x^\wsh_{1-}\, ,\x^\wsh_{2-}]
\Zp^\wsh[\x^\wsh_{i-} \,,\x^\wsh_{i+} ]
\Zp^{>}[\x^\wsh_{1+}\, ,\x^\wsh_{2+}] \, ,
\ee
with $\Zp^{<}$ ($\Zp^{>}$) the string partition function restricted to $\z<\z_\wsh$ ($\z>\z_\wsh$) while $\Zp^\wsh$ is the partition function in the neighborhood of $\z=\z_\wsh$.

Since the action is 
quadratic, the partition function  
 is given by the classical action up to a constant independent of the endpoint 
\be
\label{Zwsh}
\Zp^\wsh[\x^\wsh_{i-} \,,\x^\wsh_{i+} ]\ &\propto & \
e^{
	\ \ -i\frac{1}{2}  \int \frac{\dd\w}{2\pi} 
	\left[
	\Tenp \x^\wsh_{1+}(-\w) \del_\z \X_{1+}(\w,\z_+)
	-
	\Tenm\x^\wsh_{1-}(-\w) \del_\z \X_{1-}(\w,\z_-)	
	\right]
     } 
     \nonumber
     \\
     && \!\!\!
     \times\, e^{
	i\frac{1}{2}\int \frac{\dd\w}{2\pi}
	\left[
	\Tenp\x^\wsh_{2+}(-\w) \del_\z \X_{2+}(\w,\z_+)
	-
	\Tenm\x^\wsh_{2-}(-\w) \del_\z \X_{2-}(\w,\z_-)	
	\right]
     } 
     \, .
\ee 
For sufficiently small $\epsilon$  we have $\Ten(\z_-)\approx -\Ten(\z_+)=\Tenwsh$.

The classical solution  \Eq{clsl} may be expressed in terms of the $\x^{\wsh}_{i\pm}$.
\be
a_i(\w)&=&\frac{1}{e^{\delta_i  \w/2T} -1} \frac{\x^\wsh_{i+}(\w)-\x^\wsh_{i-}(\w)}{F^{\wsh*}_\w} \, ,
\\
b_i(\w)&=&\frac{1}{e^{\delta_i  \w/2T} -1} \frac{e^{\delta_i  \w/2T} \x^\wsh_{i-}(\w)-\x^\wsh_{i+}(\w)}{F^\wsh_\w} \, ,
\ee
where, again, $F_\w(\z_+)\approx F_\w(\z_-)=  F^\wsh_\w$

Following \cite{Son:2009vu} we introduce the ``$ra$'' basis
\be
\X_r(\w,\z)=\frac{1}{2}\left(\X_1(\w,\z) + \X_2(\w,\z)\right) \, , \quad \quad 
\X_a(\w,\z)=\left(\X_1(\w,\z) - \X_2(\w,\z)\right) \, .
\ee
In this  basis, the exponent of \Eq{Zwsh} is given by
\be
\label{seff}
iS^{eff}&=&i\frac{1}{2}\Tenwsh\int \frac{\dd\w}{2\pi} \times
\\
&&
\quad 
\left[
 2 i{\rm Im} \left\{ \frac{\del_\z F_\w^\wsh}{F_\w^\wsh} \right\} \frac{e^{\w/2T}+1}{e^{\w/2T}-1} 
        \left(
        \x^\wsh_{r-}(-\w)-\x^\wsh_{r+}(-\w)
        \right)
         \left(
        \x^\wsh_{r-}(\w)-\x^\wsh_{r+}(\w)
        \right)
        \right.    
\nonumber
\\
&&
\quad 
+\frac{i}{2} {\rm Im} \left\{ \frac{\del_\z F^\wsh_\w}{F^\wsh_\w} \right\} \frac{e^{\w/2T}+1}{e^{\w/2T}-1} 
        \left(
        \x^\wsh_{a-}(-\w)-\x^\wsh_{a+}(-\w)
        \right)
         \left(
        \x^\wsh_{a-}(\w)-\x^\wsh_{a+}(\w)
        \right)    
\nonumber
        \\
        &&
        \quad
+\x_{a-}^\wsh(-\w)\left(
		\x_{r-}^\wsh(\w)\left(\frac{\del_\z F^\wsh_\w}{F^\wsh_\w}+\frac{\del F^{\wsh*}_\w}{F^{\wsh*}_\w}\right)
		+\x_{r+}^\wsh(\w)\left(\frac{\del_\z F^\wsh_\w}{F^\wsh_\w}-\frac{\del F^{\wsh*}_\w}{F^{\wsh*}_\w}\right)
		\right)    
\nonumber
		        \\
        &&
        \quad
        \left . 
-\x_{a+}^\wsh(-\w)\left(
		\x_{r+}^\wsh(\w)\left(\frac{\del_\z F^\wsh_\w}{F^\wsh_\w}+\frac{\del F^{\wsh*}_\w}{F^{\wsh*}_\w}\right)
		+\x_{r-}^\wsh(\w)\left(\frac{\del_\z F^\wsh_\w}{F^\wsh_\w}-\frac{\del F^{\wsh*}_\w}{F^{\wsh*}_w}\right)
		\right) 
		\right]   \, ,
\nonumber
\ee
where  \Eq{tencont} has been used.
The action can be  simplified  further since
\be
\Tenwsh\frac{\del_\z F^\wsh_\w}{F^\wsh_\w}=i\w \eta \, , \quad \eta=\frac{1}{2}\sqrt{\lambda}\pi T^2
\, ,
\ee
as it can be easily seen from \Eq{Fdef} and \Eq{Tensiondef}.

The partition function $\Zp^>[\x^\wsh_{i+}] $ depends on the fluctuations at $\z>\z^\wsh$ and, in
particular at the AdS horizon $\z^h=(1-\epsilon)\sqrt{\gamma}/\pi T$. 
\be
\Zp^>[\x^\wsh_{1+},\x^\wsh_{2+}]= 
\int \D \x_1^h(\w) \x_2^h(\w) \Zp^>_b[\x^\wsh_{1+},\x^\wsh_{2+},\x^h_{1},\x^h_{2}] 
\Zp^h 
\left[\x^h_1, \x^h_2
\right] \, ,
\ee
where we have made explicit the dependence on the coordinates at the horizon 
$\x^h_i(\w)=\X_i(\w,\z^h)$.
Since we want to describe the string
dynamics at all scales causally connected with the boundary, we perform the matching between the 
fluctuations on the left and right sides of the Kruskal plane, \Fig{kruskal}, at $\z^h$.   As  in  \Eq{Zwsh}
the partition function
at the horizon is given by the classical action
\be
\label{zhcl}
\Zp^h\left[\x^h_1, \x^h_2
\right] \propto e^{-i\frac{\Ten(\z^h)}{2} \int \frac{d\w}{2\pi}
		 \left[
		 	\x^h_2(\w) \del_\z \X_2(\w,\z^h)-\x^h_1(\w) \del_\z \X_1(\w,\z^h)
		\right]		
							}	
							\,. 
\ee
An analysis similar to the one performed at $\z^\wsh$ must be performed at  $\z^h$.  The classical solution 
\Eq{clsl} close to the horizon sets 
\be
\label{thbc}
\X_{a}(\w,\z^h)=0\, .
\ee
Thus, the effective action $\Zp^h$  forces the string fluctuations to fulfill \Eq{thbc}.   To
show this, we perform the matching slightly above the horizon in the left universe and identify 
\be
\x^h_2(\omega)=\X_2(\z^h+\epsilon,\omega) \, ,
\ee
and we will take  $\epsilon\rightarrow0$ at the end\footnote{This procedure avoids the complication that in  \Eq{clsl} 
the solutions in the left and right universe above the world-sheet horizon are proportional.
}. To leading order in $\epsilon$ we find 
\be
\label{ah}
a(\w)&=&\frac{x_a(\omega)}{2 \epsilon \del_\z F^{*}_\w(\z^h)} \,  ,
\\
\label{bh}
b(\w)&=&\frac{x_a(\omega)}{2 \epsilon \del_\z F_\w(\z^h)}  \, .
\ee

Substituting the classical solution \Eq{clsl}, \Eq{ah} and \Eq{bh} in \Eq{zhcl} and taking the the small $\epsilon$ limit 
we obtain 
\be
\Zp^h\left[\x^h_1, \x^h_2
\right] &=&\lim_{\epsilon\rightarrow0} \exp\left\{
-i \Ten(\z^h)\int \frac{d\w}{2\pi}  \frac{\x^h_a(-\w) \x^h_a(\w) }{4 \epsilon}
\right\} \,  ,
\\
\label{Xazero}
&=&\N \delta\left(\x^h_a(\w) \right) \, ,
\ee
with $\N$ a divergent constant (independent of the string coordinates) which can be absorbed into the normalization of the partition function. Note also that, unlike the $v=0$ case \cite{Son:2009vu}, the partition function 
does not depend on the 
string coordinates at $\z^h$.  The horizon effective action \Eq{Xazero} breaks the symmetry between the
$r$ and $a$ coordinates in the string action.

The partition function in the bulk is also conveniently expressed in the $r a $ basis. Integrating the
classical action by parts, we find
\be
\Zp^< 
			\left[
			\x^\wsh_{1-}\, , \x^\wsh_{2-}
  		   	\right]
&=& 
	\int \D \x^\b_a (\w) \D \x^\b_r (\w) \,
	 \int^{\X_i(\w,\z_-)=\x^\wsh_{i-}}_{\X_i(\w,\bepsilon)=\x^\b_i}  \Ds \X_a(\w,\z)  \Ds \X_r(\w,\z)
	\\
	&&
	\quad \
	\left[\,
e^{i  \Tenb \int \frac{\dd\w}{2\pi}\, \x^\b_a(-\w) \del_\z \X_r(\w,\bepsilon)} \,
e^{-i \Tenm \int \frac{\dd\w}{2\pi}\, \x^\wsh_{a-}(-\w) \del_\z \X_r(\w,\z_-)}
	\right . \times
\nonumber
\\
&&
\quad \quad
\left . 
e^{i  \int \frac{\dd\w}{2\pi} \int^{\z_-}_\bepsilon \dd\z\,
      \X_a(-\w,\z) 
      \left[\del_\z \left(\Ten(\z) \del_\z \X_r(\w,\z)\right)+\frac{m \w^2}{\pi T \z^2f(\z)}\X_r(\w,\z)\right]
    }
    \right]
    \nonumber \, ,
    \\
\Zp^> 
			\left[
			\x^\wsh_{1+}\, , \x^\wsh_{2+}
  		   	\right] 
&=&
	 \int_{\X_i(\w,\z_+)=\x^\wsh_{i+}}  \Ds \X_a(\w,\z)  \Ds \X_r(\w,\z) \nonumber \\
	 && \quad \
\left[\,
e^{i  \Tenp \int \frac{\dd\w}{2\pi}\, \x^\wsh_{a+}(-\w) \del_\z \X_r(\w,\z_+)} \ \times
\right .
\nonumber
\\
&&
\quad \quad
\left . 
e^{i  \int \frac{\dd\w}{2\pi} \int^{\z^h}_{\z_+} \dd\z\,
      \X_a(-\w,\z) 
      \left[\del_\z \left(\Ten(\z) \del_\z \X_r(\w,\z)\right)+\frac{m \w^2}{\pi T \z^2f(\z)}\X_r(\w,\z)\right]
    }
    \right] \, ,
\ee
where we have introduced the notation $\Tenb=\Ten(\z=\bepsilon)$.
In writing $\Zp^> 
			\left[
			\x^\wsh_{1+}\, , \x^\wsh_{2+}
  		   	\right] $ we have used \Eq{thbc}.
			
\subsection{Stochastic String Motion.}
The ``$a$'' coordinates in the partition function \Eq{Zwsh} can be integrated out, after introducing a random
force term in \Eq{seff}    
\be
&& e^{- \frac{1}{2}\int \frac {\dd\w}{2\pi}\, \frac{\Tenwsh}{2} {\rm Im} \left\{\frac{\del_\z F^\wsh_\w}{F^\wsh_\w} \right\} \frac{e^{\w/2T}+1}{e^{\w/2T}-1} 
        \left(
        \x^\wsh_{a-}(-\w)-\x^\wsh_{a+}(-\w)
        \right)
         \left(
        \x^\wsh_{a-}(\w)-\x^\wsh_{a+}(\w)
        \right)
        } 
        \\
        &&=
        \int  \D\hxi^\wsh
        e^{i \int \frac{\dd\w}{2\pi} 
                \left(
        		\x^\wsh_{a-}(-\w)-\x^\wsh_{a+}(-\w)
        	     \right)
        		\hxi^\wsh(\w) }
     e^{	-\frac{1}{2}
     		 \int \frac {\dd\w}{2\pi}
		 		 \frac{2}
		         { \Tenwsh }  
			 {\rm Im}
		         \left\{\frac{F^\wsh_\w}{\del_\z F^\wsh_\w} \right\}
		          \frac{e^{\w/2T}-1} {e^{\w/2T}+1}
			\hxi^\wsh(-\w)\hxi^\wsh(\w)
          }
          \nonumber
          \, .
\ee 

The quadratic term in $\x^\wsh_{r\pm}$ in  \Eq{seff} can be understood as a discontinuity in the 
average string position across the world-sheet horizon
\be
\label{disc}
\x^\wsh_{r+}=\x^\wsh_{r-} + \Delta\, .
\ee
The discontinuity $\Delta$ is random, with a Gaussian distribution given by \Eq{seff}.
This discontinuity is natural, since the correlation between the two sides of the horizon
can be viewed as a tunneling process across a barrier.
Close to the horizon, the fluctuations are large and 
the procedure of analytically continuing across the horizon is similar to the matching of
WKB solutions in quantum mechanics in the vicinity of a turning point.

Integrating over the ``$a$'' coordinates, the partition function is expressed as a product of $\delta-$functions.
\be
\label{fspf}
\Zp&=&
			\int \D \x^\b_{r}  \, \D \x^\b_{a} \,  \D \x^\wsh_{r-} \Ds \X_r
             		\llangle
		          e^{
						i  \int \frac{\dd\w}{2\pi}
						 \, \Tenb \x^b_{a}(-\w) \del_z \X_{r}
										\left(
										\w,\bepsilon
										\right)
						}
\right.
\\
& &
\quad \quad \quad \quad 
\left.
\times
						 \delta
      \left(\del_\z \left(\Ten(\z) \del_\z \X_{ r}(\w,\z)\right)+\frac{m \w^2}{\pi T \z^2f(\z)}\X_{ r}(\w,\z)\right)
      \right.
      \nonumber
      \\
      &&
      \quad \quad \quad       \quad 
      \left.
      \times \delta
      	\left(
		\Tenm\del_\z \X_{ r}(\w,\z_-) - \hxi^\wsh(\w)-
			\Tenwsh \frac{\del_\z F^\wsh_\w}{F^\wsh_\w}\left(\x^\wsh_{r-}(\w) +\Delta(\w)\right)
	\right)
	\right . 
	\nonumber
	\\
	&&
	      \quad \quad \quad       \quad 
	\left .
      \times \delta
      	\left(
		\Tenp\del_\z \X_{r}(\w,\z_+) - \hxi^\wsh(\w)-
			\Tenwsh \frac{\del_\z F^\wsh_\w}{F^\wsh_\w}\x^\wsh_{r-}(\w) 
	\right)
	\rrangle_{\hxi\, , \Delta} \, ,
\nonumber
\ee
where the subscripts $\hxi$, $\Delta$ denote average with respect to the random force and 
discontinuity variables. These are given by a gaussian distribution such that
\footnote{
Note that in the $\lambda\rightarrow \infty$ the $\left<\Delta(-\w) \Delta(\w)\right>$ vanishes and the discontinuity 
disappears. 
}
\be
\label{xicor}
\llangle \hxi^\wsh(\w) \hxi^\wsh (-\w)\rrangle&=& \frac{1}{2} \Tenwsh
{\rm Im} \left\{\frac{\del_\z F_\w^\wsh}{F^\wsh_\w} \right\}
\frac{e^{\w/2T}+1}{e^{\w/2T}-1} \, ,
\\
\label{Dcor}
\llangle \Delta(\w) \Delta (-\w)\rrangle&=& \frac{1}{2} \frac{1}{ \Tenwsh 
                                    {\rm Im} \left\{\frac{\del_\z F^\wsh_\w}{F^\wsh_\w}\right\}}
\frac{e^{\w/2T}-1}{e^{\w/2T}+1} \, .
\ee

The partition function is determined by the  classical solutions to the string equations of
motion with von Neumann boundary conditions at the boundary, the
boundary conditions at the ws-horizon
\be
\label{heqm}
\Tenm\, \del_\z \X_{r}(\w,\z_-)&=&i\w \eta  \left(
						\x^\wsh_{r-}(\w) + \Delta(\w)
						\right)
						+\hxi(\w) \, ,
\\
\label{heqp}
\Tenp\, \del_\z \X_{ r}(\w,\z_+)&=&i\w \eta 
						\x^\wsh_{r-}(\w) 
						+\hxi(\w)
\, ,
\ee
and a discontinuity at the ws-horizon 
\Eq{disc}.
The classical solution is given by
\be
\label{gensol}
\X_{r}(\w,\z)&=&
\x^\b_{r}(\w) F_\w(\z) +  \frac{{\rm Im} F_\w(\z)}{-{\rm Im} G_R(\w)} F^\wsh_\w
                \left(\hxi^\wsh (\w) + \Tenwsh \frac{\del_\z F^\wsh_\w}{F^\wsh_\w} \Delta(\w)\right)
\nonumber
\\
&&
\quad \quad \quad \quad \quad
+ \theta\left(\z \pi T-1\right) \Delta(\w) \frac{F^{*}_\w(\z)}{F^{\wsh*}_\w}  \, ,
\ee
where ${\rm Im} G_R$ has been defined in \Eq{GRdef}.

The equations of motion propagate the stochasticity at the world-sheet horizon to all 
scales. From \Eq{gensol} the bulk two point function for transverse fluctuations is given by
\be
\label{dxcorr}
\hat{G}_{T \, \sym} (\w,\z,\z') &\equiv& 
\llangle\Delta \X_{r }(\w,\z) \Delta \X_{r}(-\w,\z') \rrangle
\\
&=&
-\frac{{\rm Im} F_\w(\z) {\rm Im} F_\w(\z') }{{\rm Im} G_R(\w) }(1+2\n)
\nonumber
\\
&&
+\theta(\z\pi T-1) \frac{1}{2} \frac{{\rm Im} F_\w(\z') i  F^*_\w(\z) }{{\rm Im} G_R(\w) } \frac{e^{\w/2T}-1}{e^{\w/2T}+1}
\nonumber
\\
&&
+\theta(\z' \pi T-1) \frac{1}{2} \frac{{\rm Im} F_\w(\z) (-i)  F_\w(\z') }{{\rm Im} G_R(\w) } \frac{e^{\w/2T}-1}{e^{\w/2T}+1}
\nonumber
\\
&&
-\theta(\z' \pi T-1)\theta(\z \pi T-1)\frac{1}{2} \frac{F^*_\w(\z) F_\w(\z')}{{\rm Im} G_R(\w) }
 \frac{e^{\w/2T}-1}{e^{\w/2T}+1} \, ,
 \nonumber
\ee
where $\n=1/(e^{\w/T}-1)$ and we have subtracted the average string position
\be
\Delta \X_r(\w,\z)=\X_r(\w,\z)-\x^\b_r(\w) F_\w(\z) \, .
\ee

Finally, we may express the two point correlator in the original $t\, ,z$ coordinates. Undoing the change
of variables \Eq{chvrbl}, 
\be
\label{dxdxcor}
G_{T\, \sym}\left(t, z ; t', z' \right)
&=&\frac{1}{\sqrt{\gamma}} \int \frac{\dd\omega}{2\pi} e^{-i \omega(t-t')} e^{-i\omega(\zeta(z)-\zeta(z'))}
\hat{G}_{T\, \sym}	\left(
			\omega \sqrt{\gamma}; z \sqrt{\gamma}, z' \sqrt{\gamma}
			\right)
			\, ,
			\nonumber \\
\ee
where the function $\zeta(z)$ has been defined in \Eq{chvrbl} and relates the AdS $t$-coordinate 
to $\t$.

\subsection{Heavy Quark Partition Function\label{HQDym}}
After integrating out  $\x^\wsh_{r-}$ in the string partition function \Eq{fspf} we get the partition function for the
string boundary end point. This is the partition function for the
heavy quark. Since $\z^\b<1$ we find
\be
\label{HQpf}
\Zp^Q_T=\int \D \x^\b_{a} \D \x^\b_{r } 
						\llangle 		         
						 e^{
						i  \int \frac{\dd\w}{2\pi}\,
						 \x^\b_{ a}(-\w) 
						 \left(
						 \x^\b_{ r}(\w)
						 \Tenb \del_\z F_\w
										\left(
										\bepsilon
										\right)
						+ \hxi^\b(\w)
						\right)
						}
						\rrangle_{\hxi^\wsh\, ,\Delta} \, ,
\ee
where we have used ${\rm Im} \, G_R= -\Tenb {\rm Im} \, \del_\z F_\w (\bepsilon) $ and we have 
defined the boundary force as
\be
\label{boundaryforce}
\hxi^\b(\w)=F^\wsh_\w\left(\hxi^\wsh(\w) + \Tenwsh \frac{\del_\z F_\w^\wsh}{F_\w^\wsh} \Delta (\w)\right) \, .
\ee
It is easy to show by integration that the boundary force distribution is also Gaussian with the  
second moment given by 
\be
\llangle
\hxi^\b(\w) \hxi^\b(-\w)
\rrangle
=- \left(1+2\n\right) {\rm Im } \, G_R(\w) \, .
\ee

The first term in the exponent of \Eq{HQpf} is divergent  \cite{Son:2009vu}
\be
\label{renGr}
\Tenb \del_\z F_\w (\bepsilon)=\frac{\sqrt{\lambda}}{2\bepsilon}\w^2- G_R(\w)  \, ,
\ee
however, this divergence can be understood
as the contribution of the (large) quark mass $\Mqo=\sqrt{\lambda}/2\epsilon$ with 
$\bepsilon=\sqrt{\gamma} \epsilon$. 

Using \Eq{renGr} and undoing the change of variables \Eq{chvrbl} we find 
\be
\label{HQpf2}
\Zp^Q_T&=&\int \D x^\b_{a} \D x^\b_{r} 
						\llangle 		         
						 e^{
						i  \int \frac{\dd\omega}{2\pi}\,
						 x^\b_{a}(-\omega) 
						 \left( \gamma \Mqo \omega^2
						 x^\b_{r}(\omega)
						 -\sqrt{\gamma}G_R(\sqrt{\gamma}\omega)
						 x^\b_{ r}(\omega)
						+\xi(\omega)
						\right)
						}
						\rrangle_{\xi ,\Delta} \, ,
						\nonumber 
						\\
						&&
\ee
with $\xi(\omega)\equiv\sqrt{\gamma}\hxi^\b(\sqrt{\gamma}\omega)$.
Integrating out the $x^{b}_{a}$ coordinate we find that the average position of the quark 
follows the equation
\be
\label{beom}
 \gamma \Mqo \,\frac{\dd^2 x^\b_{r} }{\dd t^2 } + 
   \int \dd t'\, G_R\left(\frac{t-t'}{\sqrt{\gamma}}\right) x^\b_{ r}(t') 
=\xi\left(t\right) \, ,
\ee
as previously derived in  \cite{Giecold:2009cg}.

The effective equation of motion may be further clarified by studying the long time dynamics 
$\omega\rightarrow 0$ limit of \Eq{beom}. Expanding the retarded correlator to second order
\be
G_R(\w)=-i\eta \w + \frac{\Delta M}{\sqrt{\gamma}} \w^2 \, ,
\ee
with 
\be
\label{mseq}
\Delta M= \frac{\sqrt{\gamma \lambda} T}{2} \, .
\ee
Denoting $\Mq_{\rm kin}=M-\Delta M$, we obtain

\be
 \gamma \Mq_{\rm kin}\,  \frac{\dd^2 x^\b_{T\, r}}{\dd t^2} +\eta \gamma \frac{d x^\b_{T\, r}}{dt} 
=
\xi\left(t\right) \, ,
\ee
where we have recovered the subscript $T$ to denote transverse fluctuations. Note that, due to the mass shift
\Eq{mseq}, the heavy quark becomes effectively lighter as the velocity increases.  This reduction 
of  the kinetic mass coincides with the result argued in \cite{Beuf:2008ep}.

\subsection{Extension to the longitudinal fluctuations\label{lf}}
The analysis of longitudinal fluctuations around the trailing string solution is completely analogous
to the one performed above for transverse modes. Thus, we will not repeat the computation here
but we will simply state the main results. From the observation that the action is multiplied by a 
factor $\gamma^2$ a simple rule can be given to obtain the longitudinal expressions from the 
transverse ones; it is sufficient to replace $\Ten\rightarrow \gamma^2 \Ten$. 

The stochastic longitudinal fluctuations may be expressed as 
\be
\X_{L\,r}& =&\x^\b_{L\, r} F_\w (\z) + \frac{{\rm Im} F_\w(\z)}{-\gamma^2{\rm Im} G_R(\w)}
F^{\wsh}_\w
		\left(
			\xi^\wsh_L+ \gamma^2 \Tenwsh \frac{\del_\z F^\wsh_\w}{F^\wsh_\w} \Delta_L(\w)
		\right)
\nonumber
\\
&&
+
\theta(\z-1)\Delta_L(\w) \frac{\del_\z F^*_\w(\z)}{F^{\wsh*}} \, ,
\ee
with $\x^\b_{L\, r}$ is the boundary value of the fluctuation and $\xi^\wsh_L$ and $\Delta_L$ are 
random variables distributed according to the two point functions
\be
\llangle
\xi^\wsh_L(\w) \xi^\wsh_L(-\w)
\rrangle
&=&\gamma^2
\llangle
\xi^\wsh(\w) \xi^\wsh(-\w)
\rrangle \, , 
\\
\llangle
\Delta_L(\w) \Delta_L(-\w)
\rrangle
&=&\frac{1}{\gamma^2}
\llangle
\Delta(\w) \Delta(-\w)
\rrangle \, ,
\ee
with the transverse correlators defined in \Eq{xicor} and \Eq{Dcor}.

The correlation function for string fluctuations around the average value is also proportional to the 
longitudinal correlator
\be
\G_{L \, \sym} (\w, \z, \z') \equiv 
    \llangle
    \Delta \X_{L r} (\w,\z) \Delta \X_{L r} (-\w, \z')
    \rrangle 
    =\frac{1}{\gamma^2}
     \G_{T, \sym} (\w,\z,\z')
\ee

On the boundary, the stochastic string motion leads to a random longitudinal force which is related
to the random variables $\xi^\wsh_L$, $\Delta_L$ as
\be
\xi_L(\omega)=\sqrt{\gamma} F^\wsh_\w
\left(
\xi^\wsh_L(\w) + \gamma^2 \Tenwsh \frac{\del_\z F^\wsh_\w}{F^\wsh_\w} \Delta_L(\w)
\right) \, .
\ee 
The boundary force-force correlator is 
\be
\llangle \xi_L(\omega)\xi_L(-\omega)\rrangle
=-\gamma^{5/2} \left(1+2n(\sqrt{\gamma} \omega)\right)
{\rm Im} G_R (\sqrt{\gamma} \omega)
\ee

Finally, the effective equation of motion for the boundary end-point of the string is given by 
\be
\gamma^3 \Mqo \frac{\dd^2 x^\b_{L \,r}}{\dd t^2} + \gamma^2 \int \dd t'\, G_R\left(\frac{t-t'}{\sqrt{\gamma}}\right)
x^\b_{L\, r}(t') = \xi_L(t) \, ,
\ee
which in the low frequency limit is 
\be
\gamma^3 \Mq_{\rm kin}\frac{\dd^2 x^\b_{L \,r}}{\dd t^2}+ \gamma^3 \eta \frac{\dd x^\b_{L\, r}}{\dd t}
=\xi_L(t) \, .
\ee

\section{Summary and Outlook\label{sando}}
\subsection{Summary of the main results}
In this work we have analyzed the stochastic motion of the trailing string generated at the 
world-sheet horizon by integrating out the near horizon string fluctuation.
  In contrast to  the static case, the string fluctuations below the world-sheet horizon are 
causally connected to  the boundary. The integration of these modes leads to a discontinuity in the 
string across the world-sheet horizon $\Delta_\alpha$ and the generation of a random force  acting on the string above
and below the horizon $\xi^{\wsh}_\alpha$. The subindex $\alpha=L,\,T$ indicates that  the force is different for the longitudinal and transverse components.
The discontinuity $\Delta_\alpha$ should not be understood as a discontinuity of the 
wold-sheet since our description in term of small string fluctuations is only valid outside of a 
small region around the world-sheet horizon; the discontinuity  refers to the different positions 
of the string above and below the stretched horizon.
 We
have 
sketched these dynamics in \Fig{forcebalance}.

\FIGURE[t]{
\epsfig{file=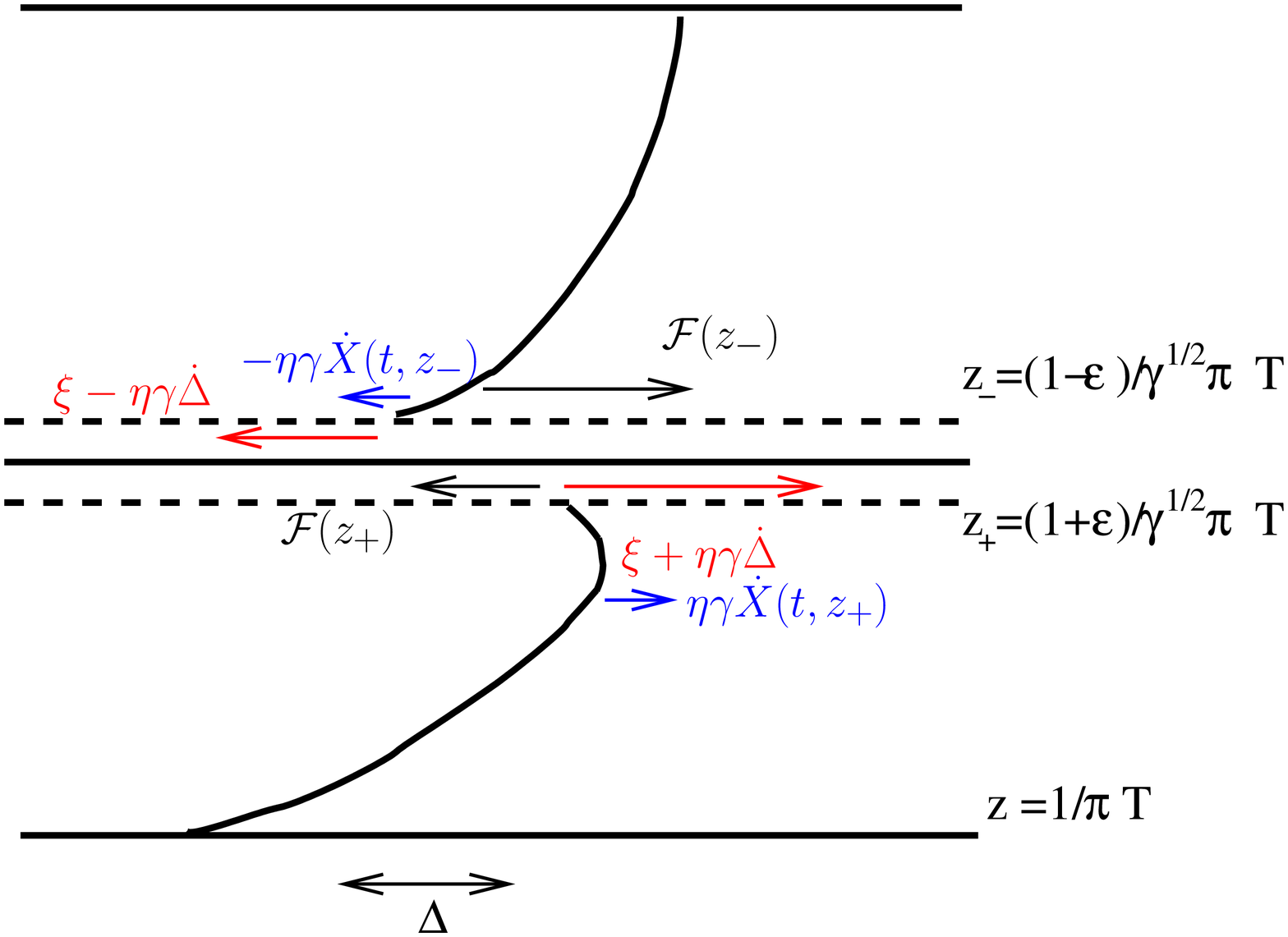,width=10cm}
\caption
{
Sketch of the string profile across the world-sheet horizon. The string fluctuation within the 
stretched horizon lead to discontinuity in string position $\Delta$ and the appearance of a random
force $\xi$ which acts on both ends of the string. The tension force across the world-sheet horizon is 
discontinuous due to the different velocities of the string endpoints. There is a balance of forces
on each side of the world-sheet horizon among the tension force, the random force and a resistive
force $\pm\eta \gamma \dot X(t, z_\pm)$ and a force which depends on the separation  of the string endpoint
$\eta \gamma \dot \Delta(t)$. Note that 
we plot the longitudinal fluctuations for clarity; there are also fluctuations transverse to the string 
motion. 
}
\label{forcebalance}
}

The computations in this work have been performed in the set of coordinates  \Eq{chvrbl}. In this 
discussion we will undo this change of variables and restore the AdS coordinates \Eq{poincare}. 
For simplicity of notation, we will only summarize the transverse fluctuations and drop the subscript $T$, since the extension to the longitudinal case is straight forward, \Sec{lf}.
From \Eq{fspf} and undoing the change of variables  \Eq{chvrbl} the forced motion of the wold sheet end points of the string is given by\footnote{
We use that $X(\omega,z)=e^{-i\omega \zeta(z)} \X(\omega \sqrt{\gamma},z\sqrt{\gamma})$ with 
$\zeta$ given in \Eq{chvrbl}. 
} 
\be
\label{eomhorizon}
\Tenpm \del_z X_{r}(t,\zpm)- \xi^\wsh(t)&=&\pm  \eta\gamma \dot \Delta
\\
\llangle
\xi^\wsh(t)
\xi^\wsh(t')
\rrangle=G^\wsh_{\sym}(t-t')
&&
\llangle
\Delta (t)
\Delta(t')
\rrangle=G_{\Delta}(t-t')
\ee
with the discontinuity variable $\Delta=X(t,\zp)-X(t,\zm)$ and the correlators are obtained from \Eq{xicor} and \Eq{Dcor}
\be
G^\wsh_{\sym}(\omega)=\frac{ 1}{2} \gamma\, \eta \omega
 \frac{e^{\omega \sqrt{\gamma}/2T}+1}{e^{\omega \sqrt{\gamma}/2T}-1} \, ,
\quad \, \quad
G_{\Delta}(\omega)=\frac{1}{2} \frac{1}{\eta{\gamma}\omega } 
 \frac{e^{\omega \sqrt{\gamma}/2T}-1}{e^{\omega \sqrt{\gamma}/2T}+1} \, .
\ee
In deriving this expression we have assumed that $\epsilon\ll 1-1/\sqrt{\gamma}$ so that 
the strip does not overlap with the world-sheet horizon. Thus, we shall not take  
$v\rightarrow 0$ below\footnote{
This is a consequence of exponentiating \Eq{chvrbl}, $e^{-i\w \t}$, which leads to two poles at $\z=1/\pi T$ and 
$z=1/\pi T$; these poles coincide at $v=0$.
 We have 
explicitly checked that we reproduce the $v=0$ limit. In this limit, the transformation \Eq{chvrbl} is trivial and the equations of motion are the same as in \cite{Son:2009vu}}


The horizon equations of motion \Eq{heqm}  and \Eq{heqp} can be interpreted as the balance of the 
random force at the horizon with the tension force, $\F$, and 
a drag like force on the string endpoints. 
\be
\F(z_-)+ \xi-\eta \gamma \dot \Delta &=& \eta \gamma \dot X(t,z_-) \, ,
\\
\F(z_+)- \xi-\eta \gamma \dot \Delta &=& -\eta \gamma \dot X(t, z_+) \, ,
\ee
where we have identified the tension  force $\F(z_\pm)=\pm\Tenpm \del_z \X(t,\z_\pm)$ and $\w=\sqrt{\gamma}\omega$ (see Appendix \ref{FB}  for a derivation of this expression). The string is discontinuous across the world-sheet horizon and the 
point above and below the horizon move with different speeds which leads to a discontinuity of the tension force. 
The equation of motion on each side of the horizon depends on the endpoint on the other side
via $\Delta$ which  might be interpreted as a force of magnitude
$\pm \eta \gamma \dot X(t,z_\pm)$ 
due to the motion of the string on the other side of the horizon. The net force introduced by the 
horizon is $\F(z_-) + \F(z_+)=\eta \gamma \dot \Delta$ and grows as the separation of the two sides of the string
grows.

The classical string equations propagate the stochastic motion on the world-sheet horizon endpoint
to all scales. 
 Of particular interest is the motion of the boundary endpoint, since it is dual to 
 the position of the heavy quark. For these, the fluctuations below the world sheet horizon may be 
integrated out leading to a second random force at the world sheet horizon. 
The effective force acting on the boundary is given by \Eq{boundaryforce}
\be
\label{effectivef}
\xi(t)=\int^t \dd t'\, F\left(\frac{t-t'}{\sqrt{\gamma}},z_-^\wsh \right)
\left(
\xi^\wsh(t') -\eta \gamma \dot \Delta(t')
\right) \, ,
\ee
where $F(t,z_-^\wsh)$ is the retarded boundary to horizon propagator \Eq{Fdef}. 
Note that the time it  takes the noise to propagate to the boundary $\sim \sqrt{\gamma}/\pi T$ 
grows with the velocity of the quark. The force correlator at the boundary is given by
\be
\label{forcecorr}
\llangle
\xi(t) \xi(t')
\rrangle= G^{v=0}_{\sym} \left(\frac{t-t'}{\sqrt \gamma }\right) \, ,
\ee
with the zero velocity symmetrized correlator
\be
G^{v=0}_{\sym}(\omega)=-(1+2 n) {\rm Im\,} G_R(\omega) \, .
\ee
From \Eq{forcecorr} is easy to conclude that the noise correlation time $\tau_C$ grows with the
velocity of the quark
\be
\label{taucdef}
\tau_C\sim\frac{\sqrt{\gamma}}{\pi T}
\ee

The combination of the random force \Eq{effectivef}  and 
von Newmann boundary conditions at the boundary leads to the equation of motion for the boundary end point
\Eq{beom}, which in the low frequency limit may
be expressed as
\be
\label{eomfinal}
\frac{d\bp}{dt}={\bf \E} -\mu {\bp} + \xi(t) \, ,
\ee
with $\mu=\eta/M_{\rm kin}$. This equation describes  the forced motion of the  particle in a dissipative medium with the random force correlators 
\be
\llangle
\xi_T(t)\xi_T(t')
\rrangle&=& \sqrt{\gamma}\kappa  \delta(t-t')\, ,
\\
\llangle
\xi_L(t)\xi_L(t')
\rrangle&=& \gamma^{5/2}\kappa  \delta(t-t') \, ,
\\
\kappa&=&\sqrt{\lambda}\pi T^3 \, .
\ee
These equations have been previously derived in \cite{Giecold:2009cg}.

As in  \cite{Son:2009vu}, the string position at an arbitrary point $z$ may be expressed as  
the reaction of the  in-falling string to  the motion at the boundary $x_0(t)$ plus a random piece. 
\be
\label{stringXoft}
X(t,z)=\int \dd t' \, F^v\left(t-t',z\right) x_0(t') + \Delta X(t,z) \, .
\ee 
where the in-falling function $F^v(\omega,z)=e^{-i\omega \zeta(z)} F(\sqrt{\gamma} \omega,\sqrt{\gamma}z)$ and $\zeta(z)$ is given in \Eq{chvrbl}.
The random function $\Delta X(t,z)$ vanishes at the boundary and it is given by the distribution
\be
\llangle \Delta X(t,z) \Delta X(t',z')
\rrangle= G_{ \sym} (t,z; t', z') \, ,
\ee
with the explicit expression for $G_{ \sym} (t,z; t', z') $ given by \Eq{dxdxcor}.

\subsection{The low frequency limit and  the physical picture}
To further clarify the dynamics of the string, we consider its motion  over a long time interval
$\tau$ such that  
\be
\tau \gg \tau_C\sim \frac{\sqrt{\gamma}}{\pi T} \, ,
\ee
{\it i. e.}, much larger than the correlation time of the noise. Over this time interval, the dynamics
are well approximated by the low frequency limit of the string motion. 

We study the reaction of the string to  a small fluctuation $x_0(t)$ on the quark motion over 
the stationary trajectory  induced by the external electric field.  This fluctuation is relaxed in a time
given by the inverse drag coefficient $\mu$
\be
\label{taurdef}
\tau_R \sim \frac{M_{\rm kin}}{\sqrt{\lambda}T^2}
\ee
which, for sufficiently heavy quarks, is large compared to $\tau_C$. We will demand 
the time interval $\tau\ll\tau_R$ so that the velocity can be considered as stationary.
In this time interval, the average motion of the string is obtained from the low frequency expansion 
of \Eq{stringXoft}
\be
\label{averageX}
\llangle
X(t,z)
\rrangle=x_0(t) + \dot x_0(t) \Delta X_{TS}(z) \,,
\ee
with the trailing string profile given by  \Eq{tsprfl}.
Thus, the reaction of the string to
the perturbation leads to a trailing string solution with a velocity given by 
${\bf v}= v {\hat x}+\dot {\bf x}_0(t)$.

\FIGURE[t]{
\epsfig{file=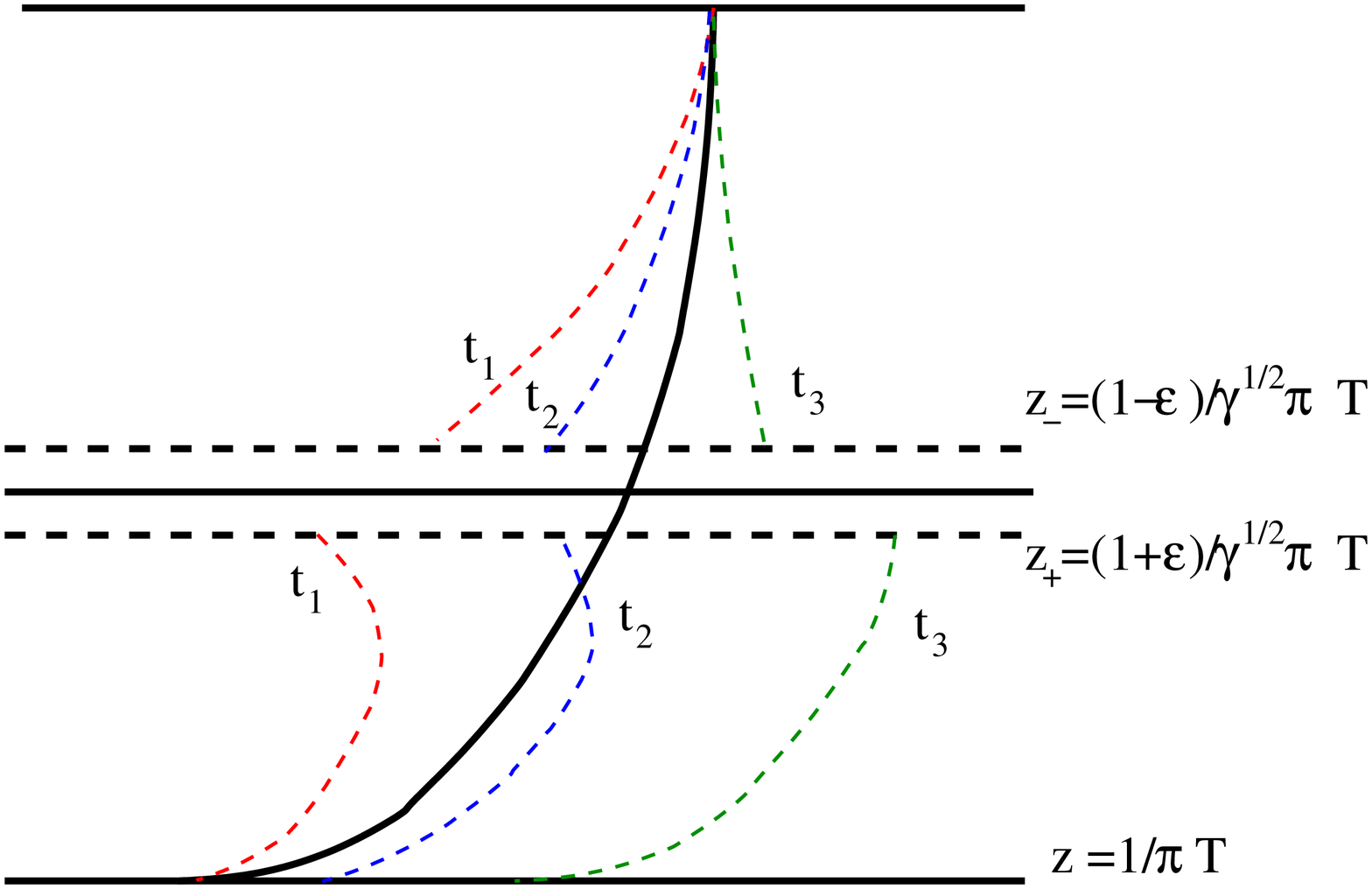,width=10cm}
\caption
{
Schematic view of the string fluctuations induced by the world-sheet horizon at three different times.  
On average the string
is given by the trailing string solution \Eq{dssol} with the velocity given by the motion of the boundary
end point. At any given time, the string deviates from the average by a drag-like string but with a 
characteristic higher effective temperature $\sqrt{\gamma}T$ and with a random amplitude $-\xi/\eta$.
This stochastic ensemble of stings generates a random force of the boundary $\xi$. 
The string above and below the horizon are separated by a random variable $\Delta$.  As in 
\Fig{forcebalance} we only plot the longitudinal fluctuations for clarity. 
}
\label{trailingflipflop}
}

On top of the average value \Eq{averageX}, there is a random string profile which is given in 
\Eq{gensol} which in the low frequency limit is\footnote{
Note that the discontinuity of the derivative in \Eq{eomhorizon} vanishes at $\omega=0$
}
\be
\\
\nonumber
\Delta X(t,z)&=&-\frac{\xi(t)}{\eta} \frac{1}{ 2\pi \sqrt{\gamma} T} 
\left(
\tan^{-1}\left(z \pi \sqrt{\gamma} T\right)  + \frac{1}{2} \ln
       \left|
       \frac{1-z \pi \sqrt{\gamma} T}
               {1+z \pi \sqrt{\gamma} T}
               \right|
\right) 
+\theta\left(\sqrt{\gamma} \pi T z-1\right)
\Delta(t) \, ,
\ee
\be
\llangle\xi(t) \xi(t')\rrangle= 2 \eta \sqrt{\gamma}T \delta(t-t') \,, \quad
\llangle\Delta(t) \Delta(t')\rrangle= \frac{1}{8 \eta \sqrt{\gamma}T } \delta(t-t') \, ,
\ee
This result can be interpreted as follows: the random force $\xi$ generates an ensemble of strings 
which fluctuate around the average string  \Eq{averageX}. Each of the strings in the ensemble
have the same $z$-functional form which corresponds to the trailing string profile but with a higher characteristic
temperature, $\sqrt{\gamma} T$ which is the position of the world-sheet horizon.   As the string goes
across the world sheet horizon, there is a discontinuity in the position $\Delta$ which can be 
understood as an effect of tunneling through a barrier. 
 This barrier is due to the fact that at the world-sheet horizon
the average velocity of the string is larger than the local longitudinal speed of light. 

Finally, let us remark that turning off the electric field and  in the very long time limit $\tau\gg \tau_R$ the momentum of the quark is 
relaxed and the velocity tends to its (small) equilibrium value. Thus, $\gamma\rightarrow0$ and the 
long time dynamics are described by the static string \cite{Son:2009vu}.

\subsection{Limits on  the validity of the approach\label{limitonvalidity}}
The results in this paper are based on the assumption that the heavy quark maintains an approximately
constant speed during a time which is large compared to all medium scales, so that the 
 dynamics of the quark are controlled by medium averages.  
Since the force correlation time grows with the quark velocity, \Eq{taucdef}, the requirement that 
the $\tau_C \ll \tau_R$  demands that
\be
\label{gcritical}
\sqrt{\gamma \lambda} \ll \frac{M_{\rm kin}}{T} \, .
\ee
 This velocity  limit
coincides parametrically with the one derived 
in \cite{CasalderreySolana:2007qw} by determining the maximum value
of the external electric field that the a D7 brane can support. The observation above provides a
more physical interpretation for this velocity constraint: at the critical value of $\gamma$  
the correlation time between the random forces acting on the quark becomes comparable 
to the motion time scale; thus, the structure of the interaction cannot be neglected.

We also note that there is a different constraint on the velocity coming from the fact that the kinetic 
mass of the quark must be large in order for the relaxation time to be large compared to the thermal
wavelength. As a consequence,  the medium modified mass 
$\Delta M$
should
be small compared to the vacuum mass $\Mqo$, $\Delta M\ll \Mqo$. Remarkably, this consideration
leads (parametrically) to the same critical $\gamma$ factor \Eq{gcritical}.  From the field theory side it is
not clear to us why these two constraints should coincide. 

We would also like to understand this limit on $\gamma$ from the point of view of the dual gauge 
theory.  The long correlation time  signals that the fluctuations due to normalizable
modes become large. Indeed, the assumption that  the normalizable modes are suppressed
is set by the condition that their classical action is large (and thus, their spontaneous fluctuations are unlikely).  Taking the average over the normalizable fluctuations of
\Eq{S2def} we obtain
\be
\llangle S_T[\Delta X] \rrangle \propto \frac{\T}{\sqrt{\gamma}} T \, , 
\ee
where $\T$ is the total time of 
observation\footnote{In $hat$ coordinates the action scales as $\hat \T=\T/\sqrt{\gamma}$.}.
 To derive this expression we have used that, in $hat$ 
coordinates, the only dimensionful quantity in the solution is $T$. Since the fluctuations are 
suppressed by $\sqrt{\lambda} $, the factor  $\sqrt{\lambda}$ in front of the action cancels.  
Demanding that the action is large in the typical time scale for the motion of the 
quark,
 $\T=\tau_R$ leads
to
\be
\frac{1}{\sqrt{\gamma}} \frac{MT}{\sqrt{\lambda} T^2}  \gg 1 \,,
\ee 
which coincides with the limit  \Eq{gcritical}.  Above the critical velocity, 
the emission of excited string modes cannot be neglected in the description of the heavy quark dynamics
\footnote{This provides a natural explanation to why the action of the string becomes imaginary for velocities larger than the 
critical one  \cite{Liu:2006he} since new channels, such as the emission of excited string states, appear. }.

To conclude this discussion, we would like to address what is the typical size of the 
stretched horizon  $\bepsilon$. The full Nambu-Goto action for transverse
fluctuations
\be
\label{ngnl}
S_{NG}=-\frac{1}{2\pi\lssq} \int \dd\t \, \dd\z \, \frac{L^2}{\z^2}\sqrt{1+f(\z)(\del_\z \X)^2 -\frac{1}{f(\z)} (\del_\t \X)^2 } \, .
\ee
The implicit assumption on the expansion performed in \Eq{S2def} is that the fluctuations remain 
small. However, the solution to the small fluctuation problem diverges logarithmically close to the 
horizon.  Thus, the last term inside the square root in \Eq{ngnl} grows faster than linearly sufficiently close 
to the horizon. Then, we must impose
\be
\frac{1}{\bepsilon}\llangle\left(\del_t \X (\t,\bepsilon)\right)^2\rrangle \ll 1 \, . 
\ee
Since for the typical fluctuation $\w\sim T$  and since the correlation 
function \Eq{dxcorr} is inversely proportional to $\eta$ we obtain the condition
\be
\frac{ \epsilon}{\ln^2  \sqrt{\gamma }\epsilon } \gg \frac{1}{\sqrt{\gamma\lambda}} \, ,
\ee
where we have used $\bepsilon=\sqrt{\gamma}\epsilon$. Thus, as long as the coupling $\lambda$
is large, the width of the stretched horizon can be small compared to temperature. As it also decreases with $\gamma$ it is possible to separate the  world-sheet horizon from the AdS horizon as long as
the velocity is not too small $v\gg 1/\lambda^{1/4}$.

\subsection{Outlook\label{outlook}}
As we have seen, the string fluctuations induced by the world-sheet horizon lead to the stochastic motion of the quark at the boundary. Since only the fluctuations above this horizon are causally 
connected to the quark, this correspondence between sting and quark fluctuations probes only the 
region $\z<\pi T$. However, as we have stressed, one of the interesting features of the trailing string is that the  effective horizon is outside the AdS event horizon.  As a consequence, the fluctuations below
the world-sheet horizon are causally connected to the boundary and they are reflected in the 
boundary theory via the correlation of  fields  associated to the quark. The strength of 
these correlations is, in essence, proportional to the correlation function \Eq{dxdxcor} which we rewrite here
for longitudinal fluctuations in the low frequency limit 
\be
\label{dxcorllf}
G_{L\, \sym} (\omega, z, z')&=&\frac{1}{\gamma^{5/2}} 
	\left(
	\frac{2}{\eta T} \Delta X_{TS}\left(\sqrt{\gamma}z\right)\Delta X_{TS}\left(\sqrt{\gamma}z'\right)
	\right . \nonumber
	\\
	&&
	\quad \quad\quad 
	\left.
	+\frac{1}{8 \eta T }\theta(z\sqrt{\gamma}\pi T-1)\theta(z'\sqrt{\gamma}\pi T-1)
	\right) \, .
\ee

Expression \Eq{dxcorllf} shows a very strong suppression of the fluctuations with the quark velocity. 
However, the correlation of associated fields do not need to show such a large suppression. 
To illustrate this point, we discuss the computation  of the stress tensor associated to the quark.  The
calculation proceeds by solving the back-reaction of the AdS metric to the string dual to the quark
\cite{Friess:2006fk,Chesler:2007an}. 
The small deviations from the AdS metric are sourced by the (five dimensional) stress tensor of the 
string which is given by
\be
\label{Tgen}
\T^{M N}(t,{\bf k})=-\frac{1}{2\pi\lssq} \sqrt{\frac{h}{g}}\del X^M \del X^N e^{-i v k_x ( t+ \Delta X_{TS}(z))} 
e^{-i k_x X_L(t,z) - i {\bf k_\perp} {\bf X}_T(t,z)} \, ,
\ee
where $\T^{M N}$  is the 5-dimensional stress tensor
(which is different from the stress tensor in the gauge theory).

The stress tensor \Eq{Tgen} depends on the fluctuations not only through the 
exponent but also through the dependence on the coordinates. Since we have restricted our analysis
to the small perturbation regime, we  expand  \Eq{Tgen} to leading order in $X_L\,, X_T$.  To this accuracy, the fluctuations do not change the average value of the stress tensor, which is given 
by the trailing string, but they lead to a non vanishing correlator. As an example, the $\T^{00}$
correlator at small momentum ${\bf k\rightarrow} 0$ is given by 
\be
\llangle \Delta \T_{00} \Delta \T_{00} \rrangle &=&\gamma^6\left(\frac{\sqrt{\lambda}}{2\pi} \frac{z}{L^5}\right)^2  
\left[A(z) A(z') \del_{z} \del_{z'}\llangle \Delta X_L(t,z)  \Delta  X_L(t',z') \rrangle  +\nonumber
       \right.
\\      
&&   
\quad \quad \quad \quad \quad   B(z) B(z') \del_{t} \del_{t'} \llangle  \Delta X_L(t,z)  \Delta X_L(t',z') \rrangle +
\nonumber
       \\
       &&
\quad \quad \quad \quad \quad          A(z) B(z')\del_{z} \del_{t'}  \llangle  \Delta X_L(t,z)  \Delta  X_L(t',z') \rrangle        +\nonumber \\
       &&
       \left. 
\quad \quad \quad \quad \quad           B(z) A(z') \del_{t} \del_{z'} \llangle  \Delta X_L(t,z)  \Delta  X_L(t',z') \rrangle
       \right] \, ,
\ee
where the functions $A(z)$ and $B(z)$ are given by
\be
A(z)&=&-v\left({(\pi T z)^2(1-2 v^2 - (\pi T z)^4 /\gamma^2)}\right) \,  ,
\\
B(z)&=&\frac{v(1-(\pi T z)^4/\gamma^2)}{f} \, ,
\ee
and do not vanish in the large $\gamma$ limit. The overall $\gamma^6$ factor compensates the 
explicit $\gamma^{5/2}$ suppression of the correlator \Eq{dxcorllf} leading to a strong 
apparent enhancement $\gamma^{7/2}$ of the stress tensor correlation function. However, 
note that the $\sqrt{\gamma}$ dependence on the functional form of the
correlation function \Eq{dxcorllf} may change this scaling with the energy; its exact functional form  will be addressed
elsewhere.

\noindent {\bf Acknowledgments.} 
 The work of JCS has been supported by a Marie Curie Intra-European Fellowship of the European Community's Seventh Framework Programme under contract number (PIEF-GA-2008-220207). KK is supported in part by US-DOE grants DE-FG02-88ER40388 and DE-FG03-97ER4014.

\newpage
\appendix
\section{Green\'{}s Function Derivation of the Fluctuation Patern\label{gfapproach}}
In this appendix we provide a different derivation of the two point correlation function \Eq{dxcorr}. We 
consider the string partition function with external sources 
\be
\label{pfws}
\Zp_T [J_1 \, ,J_2]= \int \Ds \X_1\Ds\X_2 e^{i S_T[\X_1]-i S_T[\X_2] +i \int \dd\t_1 \dd\z_1 \, J_1(\t_1,\z_1) \X_1(\t_1,\z_1)-i \int \dd\t_2 \dd\z_2\, J_2(\t_2,\z_2) \X_2(\t_2,\z_2)} \, .
\nonumber \\
\ee
Since in \Eq{dxcorr} the boundary value of the fluctuation $\Delta X$ is zero, we impose this 
condition  \Eq{pfws}.

The connected correlation function is given by 
\be
\label{gfdef}
i \G_{ ij} (t-t', \z, \z')=
(-1)^{(i+j)}
\left .
\frac{1}{i^2}\frac{\del^2 \ln \Zp_T}{\delta J_i (\t,\z) \delta J_j (\t',\z')}
\right |_{J_i=0} \, .
\ee

Under the presence of the source, the classical solution to the string equations of motion is 
given by
\be
\del_\z 
	\left(
	\Ten(\z) \del_\z \X_i(\w,\z)
	\right) 
	+ 
	\frac{m\w^2}{\pi T \z^2 f(\z)} \X_i(\w,\z) + J_i(\w,\z)=0 \, .
\ee
The connections of the type $1,2$ solutions is performed as in the case without sources. 

The correlation function \Eq{gfdef} is then given by  the Green\'{}s function 
\cite{Son:2009vu}
\be
\G_{ij}(\w,\z,\z')=-\frac
		{
		g_<(\z_i)g_>(\z'_j) \theta(\z'_j,\z_i)+g_>(\z_i)g_<(\z'_j) \theta(\z_i,\z'_j)
		}
		{
	 \Ten(\z'_j)W(\z'_j)
		}
		\, .
\ee
with $g_<(\z)$ ($g_>(\z')$) the  solution  of the homogeneous normalizable  in the right (left)
classical string solution and $W(z)=g'_<(z) g_>(z)-g_<(z)g'_>(z)$ is the Wronskian of these solutions. The function 
$\theta(\z_1,\z'_1)=\theta(\z_1-\z'_1)$, $\theta(\z_2,\z'_2)=\theta(\z'_2-\z_2)$, 
$\theta(\z_1,\z'_2)=1$ and $\theta(\z_2,\z'_1)=0$.

Using the analytical continuation \Eq{clsl}, the solutions $g_>(\z)$, $g_<(\z')$ are given by
\be
g_<(\z)= \frac{1}{2 i} \left( F_\w(\z)-e^{\theta(1-\z) \pi\w/2}F^*_\w(\z)\right)
&\quad\quad& ({\rm right \, quadrant})
\\
g_>(\z)= \frac{1}{2 i} \left( F_\w(\z)-e^{-\pi\w}e^{\theta(1-\z)\pi\w/2}F^*_\w(\z)\right)
&\quad\quad& (  {\rm right \, quadrant}) 
\\
g_<(\z)= \frac{1}{2 i} \left( F_\w(\z)-e^{\pi\w}e^{-\theta(1-\z) \pi\w/2}F^*_\w(\z)\right)
&\quad\quad& ({\rm left \, quadrant}) 
\\
g_>(\z)= \frac{1}{2 i} \left( F_\w(\z)-e^{-\theta(1-\z) \pi\w/2}F^*_\w(\z)\right)
&\quad\quad& (  {\rm left \, quadrant})  \, , 
\ee

After introducing the $r\,a$ basis, the symmetrized correlator is given by
\be
\G_{ \sym}=\frac{i}{4} \left(
						\G_{ 11}+\G_{ 22}+\G_{ 12}+\G_{ 21} 
					   \right) \, .
\ee
After a tedious but straight-forward computation, the symmetrized  correlator computed in this way
coincides with \Eq{dxcorr}. The extension to longitudinal fluctuations is also straight forward.

\section{Momentum Flux at the World Sheet Horizon \label{FB}}
The canonical momentum densities associated to the string are 
\be
\pi^\tau_\mu&=\frac{\delta \mathcal{L}_{NG}}{\delta \del_\tau X^\mu} &=-\frac{1}{2\pi\alpha'} g_{\mu\nu} \frac{h_{\tau \sigma} \del_\sigma X^\nu- h_{\sigma \sigma} \del_\tau X^\nu}{\sqrt{-h}} \, , 
\\
\pi^\sigma_\mu&=\frac{\delta \mathcal{L}_{NG}}{\delta \del_\sigma X^\mu}&=-\frac{1}{2\pi\alpha'} g_{\mu\nu} \frac{h_{\tau \sigma} \del_\tau X^\nu- h_{\tau \tau} \del_\sigma X^\nu}{\sqrt{-h}} \, ,
\ee
where $\mathcal{L}_{NG}$ is the Nambu-Goto lagrangian,  $h_{ab}$ is the induced metric on the world sheet and $g$ the AdS-metric. The string equations of motion are the continuity equations
\be
\label{conteq}
\del_\tau \pi^\tau_\mu + \del_\sigma \pi^\sigma_\mu =0\,.
\ee
The total energy and momentum of the string are
\be
E=-\int d\sigma \pi^0_t \, , \quad \quad p_i=\int d\sigma \pi^0_i\, .
\ee 
Using \Eq{conteq}, the change in momentum in the string between the interval $(\sigma_a, \sigma_b)$ is
\be
\frac{d p_i}{dt}= \pi^\sigma_i (\sigma_a) - \pi^\sigma_i(\sigma_b) \,.
\ee
The left hand side of this equation yields the forces acting on the string. The right hand side 
gives the tension forces at the string endpoint $\F(\sigma_a)=-\pi^\sigma_i(\sigma_a)$,
$\F(\sigma_b)=\pi^\sigma_i(\sigma_b)$.

  For the trailing string \Eq{dssol} we choose as $(\tau, \sigma)$ the AdS coordinates $(t,z)$. To leading order in 
  the fluctuations, the induced metric is 
\be
h_{tt}&=&-\frac{L^2}{\gamma^2 z^2} \left(
				1-\left(\pi T z \sqrt{\gamma}\right)^4
				\right) \, ,
				\\
h_{zz}&=&\frac{L^2}{z^2} \frac{1}{f(z)^2}
                  \left(1-
                       \left(\frac{\pi T z}
                                        {\sqrt{\gamma}
                                         }
                       \right)^4
                  \right) \, ,
                  \\
h_{tz}&=&-\frac{L^2 v^2 \left(\pi T\right)^2}{f(z)} 
\, .
\ee

The transverse force on the string end point at the world sheet horizon $z_\pm=\z_\pm/\sqrt{\gamma}$ is
\be
\pi^z_T(z_\pm)=-\left(\Tenpm \del_z X_T(t,\,z_\pm)-\eta \gamma \dot X_T(t,\,z_\pm)\right) \, .
\ee  
Using \Eq{eomhorizon} we find 
\be
\pi^z_T(z_\pm)=-\left(
			\xi(t)  \pm \eta \gamma \dot \Delta(t) -\eta \gamma \dot X_T (t,\,z_\pm)
			\right) \,.
\ee

\end{document}